\documentclass{kapproc}
\usepackage{procps,epsfig}
\normallatexbib
\newcommand{\eq}{\begin{equation}}
\newcommand{\eqx}{\end{equation}}
\newcommand{\be}{\begin{eqnarray}}
\newcommand{\ee}{\end{eqnarray}}
\newcommand{\f}[2]{\frac{#1}{#2}}

\begin{document}
\articletitle{LECTURES ON  CHIRAL DISORDER IN QCD} 
\author{Maciej A. Nowak\thanks{Lectures given at the Carg\'{e}se
Summer School, August 6-18, 2001.}}
\affil{M. Smoluchowski Institute of Physics, Jagellonian
  University, Cracow, Poland \\ Gesellschaft fuer Schwerionenforschung, Darmstadt, Germany}
\begin{abstract}
I explain the concept that light quarks diffuse
in the QCD vacuum following the spontaneous breakdown 
of chiral symmetry.  I exploit the striking analogy 
with disordered electrons in metals, identifying, among others, 
the universal regime described by random matrix theory, 
diffusive regime described by chiral perturbation theory
and the crossover between these two domains. 
\end{abstract}

\section*{Introduction}
In these lectures I review spectral aspects  of the  mechanism
of the spontaneous breakdown of the chiral symmetry in Quantum 
Chromodynamics.
 Most probably, the spontaneous breakdown of the chiral
symmetry is a collective phenomenon caused by the microscopic
disorder, in striking resemblance to the diffusive phenomena appearing in
disordered metals. 
Despite the microscopic, detailed nature of this disorder is still unknown,  the constraints arising 
from realizations of chiral symmetry in  QCD are so strong, 
that allow us to predict 
several non-trivial consequences of this phenomenon. Among most profound,
are the predictions on the spectral properties of the Dirac operator. 
As such, these are amenable to test using the lattice calculations. 

The lectures are intended  to be elementary and self-contained.
The outline of the lectures is as follows:\\
${\bullet}$ In Part 1, I introduce the basic facts on symmetries
and anomalies of the QCD, leading to fundamental low energy
constraints.\\
${\bullet}$ In Part 2, I explore the analogy between the metal
viewed as a complex quantum system  sharing universal properties with
so-called chaotic systems from one side,  and Euclidean QCD with light
quarks diffusing in disordered medium built from the lumps of the
gauge field, on the other side. In particular, I identify the relevant scales 
and determine the "diffusion constant" of the  QCD vacuum. Then I comment on the
close relation of this picture to several known descriptions  of chiral
symmetry breaking.\\
${\bullet}$ In Part 3,  I demonstrate how the  hierarchy of
spectral  scales explains the origin of the appearance of random
matrix
ensembles in QCD. Then I show  how  random matrix regime 
 regime breaks down at a certain
spectral scale (analog of Thouless energy), leading  
to various versions of chiral perturbation theory. 
I also point at  some  explicit confirmations of this picture coming from
the recent lattice calculations. Next, 
 I suggest how this picture can 
be modified/generalized in the
 presence 
of external parameters of the QCD.
 I mention  the possibility of (multi-critical) scaling.   
I conclude with an (incomplete)  list of  works on  disorder
and QCD,
 that could
be used as a guide for further reading and also as a good  starting point for the
original research.\\

\section*{Part 1}
Let me refresh  here  some low energy theorems of the  QCD, which 
will be needed for the 
second part of these lectures. 
Then I switch to the 
Euclidean regime, and I recall  the Banks-Casher relation. 
\subsection*{QCD} Quantum Chromodynamics  is a Yang Mills theory based
on the local gauge group $SU(3)_{\rm color}$. This means that all six quarks of
different flavors   interact with gluons, as well gluons interact with
themselves, with the same universal coupling $g$. Looking at
the typical masses of the quarks, we see that they cover several orders of
magnitude -- u,d,s,c,b,t, 
$\sim$ 5,7,150,1400,4400, 175000     MeV, -- respectively. 
The origin of this 
hierarchy goes beyond the strong interaction sector and is unknown. 
The fact, that three masses are less or at most comparable to the
scale of the strong interaction, and three others are well beyond the
scale, suggests a simplification, in which  we put three heavy masses equal
to infinity, and three light equal to zero. 
The heavy sector decouples  from the light, and the light one 
reveals a series of essential symmetries of the theory. 
We denote this idealization of the QCD as QCD$_{\chi}$.
First, let us note that such classical theory lacks any dimension-full parameter,
therefore it is scale invariant. Second, if we introduce the notion of left and
right-handed quarks,
\be
q_R= \f{1}{2}(1+\gamma_5) q
\,\,\,\,\,\,\,\,\,\,q_L=\f{1}{2}(1-\gamma_5)q
\ee
 we see that in the presence of massless quarks
the left and the right handed quarks interact independently with the
gluons, i.e. we have  two decoupled  chiral copies of the initial
theory:
\be
S_{QCD} =\int dt d^3x \sum_f \bar{q}_R^{(f)}iD\!\!\!\!/(A)q_R^{(f)} 
+ (R \leftrightarrow L) +S_{{\rm glue}}(A) 
\ee
 Constructing the standard set of the Noether currents
corresponding to the symmetries yields:
\be
V^{\mu}_a &=& \bar{q}\gamma^{\mu}t_a q \nonumber \\
V^{\mu}_0  &=& \bar{q}\gamma^{\mu} q \nonumber \\
A^{\mu}_a &=& \bar{q}\gamma^{\mu} \gamma_5 t_a q \nonumber \\
A^{\mu}_0 &=& \bar{q}\gamma^{\mu} \gamma_5 q \nonumber \\
J_{\rm scale}^{\mu} &=& x_{\nu} \Theta^{\mu \nu}
\ee
where $a=1,2,3,...8$ for $SU(3)_{\rm flavor}$, $t_a$ are corresponding generators
(here Gell-Mann matrices, modulo normalization ),
$\bar{q}=(\bar{u},\bar{d},\bar{s})$,  
$\Theta^{\mu \nu}$  denotes symmetric traceless energy momentum
tensor,
and we used the linear transformation for the original left and right
currents ($V=R+L$, $ A=R-L$).   

For three light flavors, we are left with a nonet of the conserved
vector currents, a  nonet of the conserved axial currents, and
conserved currents corresponding to the scale invariance. 
\be
SU(3)_A \times SU(3)_V \times U(1)_A \times U(1)_V \times [{\rm
   SCALE\,\,\,\,\,INV.]}
\ee

This classical picture is however strongly distorted at the quantum
level. 
First, QCD$_{\chi}$ has anomalies - i.e.  certain classical symmetries
are violated at the quantum level. The scale-invariance is 
broken ( 
scale anomaly), 
\be
\partial_{\mu}J^{\mu}_{\rm scale} =\Theta_{\mu}^{\mu}=\f{2\beta(g)}{g}
 \cdot \f{1}{4}  \left[ G^{\mu \nu i} G_{\mu \nu}^i \right]_{\rm renorml.}
\ee
and the appearance of beta function introduces
the scale of the strong interactions, $\Lambda_{{\rm QCD}}$ $\sim$ 200 MeV.
 
Second, the singlet  component of the axial current is also
anomalous
\be
\partial_{\mu} A^{\mu}_0 = \f{N_f}{2} \f{g^2}{8\pi^2} \tilde{G}^{\mu
  \nu i} G_{\mu \nu}^i 
\ee
A similar anomaly appears in QED, with $N_f/2 \rightarrow 1$, $g
\rightarrow e$, and non-Abelian $G^{\mu \nu i}$ replaced by Abelian electromagnetic
$F^{\mu \nu}$ tensor.
These are the all anomalies of the QCD, if no external
(e.g. electromagnetic) currents are added. The appearance of the
Gell-Mann function $\beta$ in the scale anomaly reminds  about another feature of
QCD:
the interactions between quarks and gluons get stronger at smaller
energy scales,
invalidating the perturbative calculation in low energy domain, 
contrary to the precisely opposite behavior in  QED. Quarks and gluons
interact strongly 
forming  colorless states, and 
the unraveled nature of long-wavelength  limit  of these 
interactions  is usually coded  under  the name   "confinement".

The closer look at the experimental spectrum of elementary 
excitations 
(hadrons) 
of the QCD vacuum shows that the picture is more complicated, and
another, on top of confinement,  nonperturbative phenomenon has to
take place
in QCD. 
If we look at the remaining symmetries after quantizing the QCD$_{\chi}$, 
we see that they are 
\be
SU(3)_A  \times SU(3)_V \times U(1)_V     
\label{exact}
\ee
where the last one corresponds to the baryon number
conservation ( 
$3V^{\mu}=B^{\mu}$).  The interaction preserves baryon number and 
is invariant under $SU(3)_V $  and $SU(3)_A$
symmetries.  Since the masses of up and down quarks are 
20 times smaller than the scale of the QCD, the lightest particles
in real QCD should 
show the traces of the exact symmetry (\ref{exact}).  
 This is at odds with the experiment, which shows that
the light vector-like particles differ from  axial-like particles, despite
similar flavor content - e.g. $\rho(770)$ is much lighter then axial
$a_1(1200)$.
 This asymmetry holds also for baryons of
opposite parity, (nucleon $n(940)$ versus $N(1535)  S_{11}$) and manifests
dramatically at the level of the scalars -
the lowest pseudoscalar (pion) $\pi(140)$  seems to not have a narrow chiral partner at
all  - when comparing  pion to $f_0(400-1200)$ with full width $600-1000$ MeV,
an  interpretation of $f_0$ as a {\em particle} is controversial... 
This suggests that the vacuum state of the QCD is not respecting all
the symmetries of the interaction of the QCD. This phenomenon is called 
a spontaneous breakdown of the symmetry.
Then, chirally invariant interactions of the QCD$_{\chi}$, acting on chirally
non-invariant vacuum, can indeed produce such an asymmetry in the
hadronic
spectrum.
 
The important hint comes from the observation by Vafa and Witten \cite{WV}, 
that theories with vector-like couplings (e.g. QCD) cannot break
spontaneously vector symmetries. 
We are therefore left  with the alternative: 
\be
Q_A^{a}|0> \equiv |PS^a> \neq 0
\ee
where $Q_A^a$ are  axial charges corresponding to the currents
$A_a^{\mu}$.
Vacuum state respects therefore only vector symmetries.
Since the Hamiltonian of the QCD$_{\chi}$ still commutes with all the 16
generators $Q^a_V$, $Q^a_A$
 we see that the vacuum state ( $H|0> =E_0 |0>$) is degenerated with
 the octet of the states $Q^a_A|0>$, ($a=1,2,...8$). Indeed 
\be
H |PS^a> = H Q^a_A |0> =   Q^a_A H |0> &=& E_0 Q^a_A|0> = E_0 |PS^a>\nonumber \\
\ee    
This is a basic message of the Goldstone theorem~\cite{GOLDSTONE}.
To each broken generator  of the axial current  corresponds a
massless, spinless  excitation 
corresponding to quantum number of the generator. In the case of three
flavors, 
the Goldstone theorem predicts the appearance of an octet of massless,
pseudoscalar mesons. They  correspond to massless pions (isotriplet),
kaons (two isodoublets) and
an eta (isosinglet). 

A priori, from the point of view of chiral symmetry, QCD allows (at least)  
two phases: asymmetric (Nambu-Goldstone phase), described above,  and
the symmetric one  (Wigner-Weyl phase), 
where  vacuum is respecting all symmetries, $Q^a_V |0>=Q^a_A |0>=0$.
Hence, a phase transition may happen between these two phases. 
Such phase transition can be characterized by the appearance of an
order parameter. The lowest-dimensional order parameter is the
expectation value of $\bar{q}_R q_L+ \bar{q}_L q_R$.
 It carries zero baryon number, 
is a scalar (vacuum respects space reflection), and is diagonal in
flavor. 
Note that e.g. $<\bar{u}d>$ is not invariant under rotations generated
by 
isospin matrices, therefore is not invariant under the vector,
unbroken group of the vacuum. There are infinitely many other
operators, which are order parameters, e.g. $\bar{q}_L q_R \bar{q}_R
q_L$, 
but they carry higher canonical dimension.
The lowest dimensional order parameter $<\bar{q}q>$ 
is called "quark condensate".   
It is believed that under the action of external parameters, like
e.g. 
temperature and/or density, the crossover to other phases of QCD
is possible (see lectures by Rob Pisarski and Krishna Rajagopal),
 including the phase with restored (approximate) chiral symmetry. 
It is important to mention, that despite some experimental signals
point at the restoration of the chiral symmetry~(cf.\cite{BROWNRHO}), 
there is, in my
opinion,  no 
"smoking gun" evidence for this phenomenon and most of our understanding of chiral phase
transition comes from the lattice studies (see lectures by Frithjof Karsch).
We should also remember, that  
in real QCD, the masses of u,d,s  quarks are light, but non zero,
therefore on top
of the phenomenon of spontaneous breakdown of the symmetry we have
also an explicit, albeit small,  explicit breakdown of the chiral
symmetry due to the explicit presence of mass terms $m\bar{q} q$ in
the Hamiltonian. Therefore the chiral restoration is not really a phase
transition, but rather a crossover process.    

\subsection*{Condensate, pion, GMOR}
 We still consider a QCD$_{\chi}$. Let us  come back to the  Goldstone
 theorem, and  show  that indeed the
 presence of the condensate forces the presence of massless excitations
in the spectrum. To  demonstrate this, we use the Ward identity, 
relating the correlator of  the axial SU(3) current $A^{\mu}_a$ and  pseudoscalar SU(3) current
$P^b = i\bar{q} t^b \gamma_5 q$ to scalar densities,
\be
\partial^{\mu} <0|T A_{\mu}^i(x) P^j(0)|0> = - i\delta(x)
 <0|\bar{q}[t^i,t^j]_+q|0>
\label{ward}
\ee
\underline{Exercise}. {\it Justify (\ref{ward}), making use of equal-time commutation
relations ( cf.  the lecture of Jean-Paul Blaizot).}

Since Lorentz invariance implies that Fourier transformation 
of the correlator has to take a form 
\be 
\int dx e^{ipx} <0|T A_{\mu}^i(x) P^j(0)|0> = 
p_{\mu} \Pi^{ij}(p^2)
\ee
the Ward identity says that $p^2 \Pi^{ij}(p^2) = -\delta^{ij}
<\bar{q}q>$. (Taking a derivative corresponds to multiplying the matrix
element by $ip_{\mu}$. Note that the vacuum expectation value of the second term in
$t^i t^j=\delta^{ij} + d^{ijk} t^k$ has to vanish due to the invariance under
vector transformations). 
Since the condensate (expectation value of r.h.s. of (\ref{ward})) is non-zero in the broken phase,  the spectral
function
$\Pi(p^2)$ has to contain a massless particle corresponding to the pole at
$p^2=0$.

Now we  can make use  of a powerful condition, (formulated by 't Hooft),
relating fundamental theories  with theories in which particles are
bound states of the fundamental constituents.  This condition, known
as anomaly matching condition,  states that a composite particle  has
to reproduce exactly the anomaly present in the fundamental theory. 

To see how this condition works let us consider the axial $SU(3)_A$
current
corresponding to $\pi^0$, i.e. $A^3_{\mu}=\bar{q}\gamma_{\mu}
t^3 \gamma_5 q$, with $q=(u,d,s)$ and $t^3=  {\rm  diag}(1,-1,0)$. 
This current is anomaly 
free 
 in  QCD, but if we allow quarks
couple to photons, the matrix element 
\be
 <0|\partial_{\mu} A^{\mu}_3|\gamma \gamma  > &=& <0|\partial_{\mu} \bar{u} \gamma^{\mu}
\gamma_5 u|\gamma \gamma> -    <0|\partial_{\mu} \bar{d} \gamma^{\mu}
\gamma_5 d|\gamma \gamma >  \nonumber \\
&=& \frac{\alpha}{2\pi} N_c (e_u^2 -e_d^2) <0| F \tilde{F} | \gamma \gamma>
\label{anom}
\ee
It is easy to understand the r.h.s. of (\ref{anom}).  
This is simply the difference of
two {\em 
electromagnetic} anomalous U(1) currents, corresponding to the charges
with fractions $e_u=2/3$ and $e_d=-1/3$, respectively.  
This equation represents the contribution 
from the celebrated triangle anomaly. 

From the above identity  we infer immediately the form of the matrix element    
\be 
<0| A^{\mu}_3|\gamma \gamma  > = 
-i\frac{p^{\mu}}{p^2} \frac{\alpha}{2\pi} N_c (e_u^2 -e_d^2) <0| F \tilde{F} | \gamma
\gamma> + ...
 \label{sing}
\ee
where $p$ is the momentum of both photons 
 and the dots stand for less singular terms. 
The pole here is simply the remnant of the massless quark circulating
in a triangle graph. 
If we now {\em assume} confinement,  anomaly matching condition  says
that in the hadronic (composite) world, there must exist a massless,
colorless
 object which couples to the axial current and photons and which
 reproduces exactly the pole in the anomaly above. 
Since massless baryons are excluded (cf. Vafa-Witten theorem),
 this object has to be a meson. 
Mesons fulfill the anomaly condition in an easy way. We do not need to 
circulate them inside the triangle loops like quarks, it is enough 
if there is a relation between the axial current and the pion.
This well-known relation reads 
\be
<0| A_{\mu}^a|\pi^b>=i \delta^{ab}F_{\pi} p_{\mu}
\label{PCAC}
\ee
where $F_{\pi}$ is pion decay constant. 
So anomaly matching means, that we read out the r.h.s. of (\ref{sing})
as a product of (\ref{PCAC}), massless pion propagator $-i/p^2$ and
the amplitude for $\pi^0\rightarrow \gamma \gamma$ decay (rest of the
r.h.s. of (\ref{sing}). 
The similar matching of the trace anomaly between the quark and
hadronic worlds was discussed in Dima Kharzeev lectures).
Let us come back to  Ward identity (\ref{ward}).
In the $p^2 \rightarrow 0$ limit the  correlator is fully determined by the
condensate, 
\be
 \int dx e^{ipx} <0|T A_{\mu}^i(x) P^j(0)|0> = 
-\frac{p_{\mu}}{p^2} <\bar{q}q>\delta^{ij}
\ee
Saturating the correlator with pion states, and using (\ref{PCAC}),  we read out
comparing left and right h.s,  the value of the another matrix element
\be
<0|P^j|\pi^k> =\delta^{jk} \frac{<\bar{q}q>}{F_{\pi}}
\label{pseudoscalar}
\ee
With the help of the above relations we can now prove  the 
Gell-Mann Oakes Renner relation~\cite{GMOR}.
  Introducing the small quark masses, 
we see that Dirac equation  implies
\be
\partial_{\mu} (\bar{u}(x) \gamma^{\mu} \gamma_5 d(x))= (m_u
+m_d)\bar{u}( x)i\gamma_5 d(x)
\ee
Calculating the vacuum to pion matrix element of the above 
identity , and using 
formulae (\ref{PCAC}) and (\ref{pseudoscalar}), we get
\be
 M_{\pi}^2 F_{\pi}^2= (m_u+m_d) <\bar{q}q>
\label{GOR}
\ee  
This relation shows how the gap in the Goldstone mode appears due to
the presence of the  nonzero quark mass.
 
 \subsection*{ Euclidean world}
In this subsection we present the formulae that allow the transcription
from Minkowski to Euclidean space. 
The advantages of  working in Euclidean space are two-fold:
first, several mathematical operations are well defined, 
second, the formulation is  comparable to the lattice simulations.
The following set of rules defines the transition ( l.h.s. denotes
Minkowski, r.h.s. denote Euclidean) 
\begin{itemize}
\item{Space time: $ix_0 \equiv x_4$, $x_i \equiv x_i$}
\item{Vector potentials: $A_0 \equiv iA_4$, $A_i \equiv A_i$}
\item{Gamma matrices: $\gamma_0 \equiv \gamma_4$, $\gamma_i \equiv
    i\gamma_i$, $\gamma_5 \equiv \gamma_5$} 
\item{Fermi fields: $i\bar{q} \equiv \bar{q}$,  $q  \equiv
    q$}
\item{Action: $iS \equiv -S$}
\end{itemize}
Then, Euclidean Dirac matrices obey $\gamma_{\mu} \gamma_{\nu}
+\gamma_{\nu}\gamma_{\mu} =2 \delta_{\mu \nu}$, all four matrices are
hermitian, 
as well as the Dirac operator $i\partial_{\mu}\gamma_{\mu}$.  
Finally, the  Euclidean action for QCD reads:
\be
S=\int d^4x \left[\f{1}{4} G^a_{\mu \nu}G^a_{\mu \nu}- \bar{q}(i\gamma_{\mu}D_{\mu}(A)+im)q\right] 
\ee
and partition function $Z=\exp (-S)$.

\subsection*{Banks Casher relation} 

We will now demonstrate that the chiral condensate is related to the Dirac
operator spectrum in Euclidean space-time.  
Consider a Dirac propagator in a Euclidan box  $V=L^4$ in the presence
of the gluonic background field $A_{\mu}$. From the action above, we read
that propagator is $S_F = (D\!\!\!\!/(A) +m)^{-1}$. 
We can write then:
\be
  V <0|\bar{q}(0)q(0);A|0>
&=& - {\rm TR}\,\, S_F (x,x)  = -\sum_k \frac{1}{i\lambda_k(A) +m}
\nonumber \\
& &\longrightarrow  \lim_{ V \rightarrow \infty} 
i\int d\lambda \rho(\lambda) \frac{1}{\lambda -im}
\ee
where TR denotes trace over coordinates, color, spin and Dirac
indices, 
and spectral density is defined as $\rho(\lambda)=\sum_k \delta(\lambda -\lambda_k)$
where $iD\!\!\!\!/ q_k=\lambda_k q_k$.
Now, we average the above equation over the gluonic configuration
weighted with the full QCD measure (including  standard gauge fixing etc.)
\be
 <(...)>_{\rm QCD} \equiv   \int  [dA] (...) {\rm det} (D\!\!\!\!/(A)+m)
 e^{-S_{\rm glue}}
\label{QCDweight}
\ee
As a result, we get 
\be
V<\bar{q}q>=i\int d\lambda\frac{\bar{\rho}(\lambda)}{\lambda-im}
\ee
where $\bar{\rho}(\lambda)=<\rho>_{\rm QCD}$ is the average spectral density, i.e. the density
$\rho(\lambda)$ averaged over the full QCD measure. 
As a final step, we take a chiral limit and use the relation
\be
\lim_{ m\rightarrow 0} \frac{1}{\lambda \pm im} ={\rm PV} \frac{1}{\lambda}
 \mp i\pi \delta(\lambda)
\ee
As a result, we relate  the chiral condensate to the average
spectral density around zero eigenvalues
\be
V <\bar{q} q> = -\pi \bar{\rho} (0)
\label{BC}
\ee
Note that the contribution from principle value (PV) part drops, since
due to the chiral property $[D\!\!\!\!/, \gamma_5]_+$ the 
eigenvalues come in pairs (corresponding to eigenfunctions 
$q_k$ and $\gamma_5 q_k$), so the average spectral density is an 
even function $\bar{\rho}(\lambda)=\bar{\rho}(-\lambda)$. Note that the same
property guarantees the positivity of the QCD measure, despite at the
first look $D\!\!\!\!/ +m$ seems to create complex measure.

The relation (~\ref{BC}) was first suggested  by Banks and Casher~\cite{BANKCASH}. (For the
careful discussion of the UV part of the spectrum, see
~\cite{LEUTWYLERSMILGA}).
 It is very important
that the chiral limit $m \rightarrow 0$  is taken after the
thermodynamical limit $V=L^4 \rightarrow \infty$, for otherwise the
average spectral density  would be zero. 
 The result states that in a vector like theory
with chirally symmetric spectra, the quark condensate is related to
mean spectral density at zero virtuality (i.e. at  $\lambda=0$). 

We can now ask the crucial question, what kind of mechanism can cause
spectral density to be non-vanishing at zero virtuality?
Note that the levels of  free particle  closed in the box scale like
$2\pi n/L$, with $n$ integer,  so the mean level spacing goes like $L^{-1}$, and
the average spectral density (proportional to the inverse
 of the mean level spacing) for
free particle scales 
like $L=V^{1/4}$, therefore will never be able to balance in the
thermodynamical limit the l.h.s. of (\ref{BC}) 
provided the condensate is non zero.
The only solution is that spontaneous symmetry breakdown requires
enormous accumulation of the eigenvalues in the vicinity of zero,
with the condensation rate scaling like $V$, so $\bar{\rho}(0)\sim
V$. 
This at the first look obvious fact was first emphasized and exploited
 by Leutwyler and Smilga~\cite{LEUTWYLERSMILGA}
  and  forms the cornerstone of the 
 spectral analysis of the Dirac operator.
    
We will devote the next lecture to unravel 
 the most plausible microscopic mechanism 
responsible for such a spectral behavior.

\section*{Part 2}
In this part, we outline the basic concepts of diffusion and translate
them into the Euclidean QCD language, identifying in this  way
the hierarchy of the spectral scales of the Dirac operator.  

\subsection*{ Primer on the diffusion}
The Banks-Casher relation is reminiscent of the Einstein relation
describing the conductivity $\sigma$ of degenerate gas of electrons,
\be
\sigma=2e^2  D \bar{\rho}(E_F) 
\label{einstein}
\ee
where $D$ is a diffusion constant, 
$\bar{\rho}(E_F) $ is the density of states (per spin direction)
at the Fermi surface. Is it possible that the spontaneous breakdown of 
the chiral symmetry has also  the  diffusive nature, with the Fermi
surface
replaced by the zero virtuality band?
We will show explicitly in this chapter, that this indeed is the case.

In order to prove this conjecture, we have to remind some basic facts
on the diffusion~\cite{BOOKDIFF,MONTAMBAUX}. 
The diffusion is a process, in which a typical distance covered by the
diffusing particle in a time $t$ varies as 
\be
\bar{r}^2(t) = D t
\label{smol}
\ee
where $D$ is the diffusion constant characterizing the medium.
If we consider the diffusive motion in a cube  $L^d$ of the linear size
$L$,
it is natural to define the time scale
$t_c=L^2/D$, characterizing the time during  which the particle can
probe
the whole system. 
The energy scale corresponding to this time, known as a Thouless
energy~\cite{THOULESS} (in units where $h=1$)
is
$E_c=1/t_c =D/L^2$. 
For times shorter than $t_c$, the diffusing particle can probe only
part of the volume. For even shorter times (shorter then the time
of mean free path between the "dirt" causing the diffusion, $t_e$) 
the diffusion concept is meaningless. 
On the other side, for very large times ($t_H\sim 1/\Delta$), 
the diffusive particle will
always leave the volume (see Edmond Iancu's lectures). When such times
start corresponding to the inverse of mean quantum spacing
of the quantum mechanical levels, the classical concept of the
diffusion 
also becomes meaningless. 
The above hierarchy of scales could be pictured by a cartoon, 
where we introduced also the names of four
different regimes using the terminology borrowed 
from the mesoscopic physics~\cite{BOOKDIFF,MONTAMBAUX}:
Quantum ($t>t_H$), Ergodic (universal) ( $t_H <t<t_c$), Diffusive (
$t_c<t < t_e $) and Ballistic ($t<t_e$).

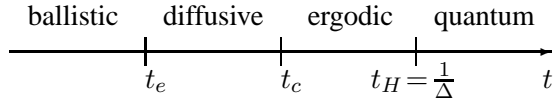
\begin{figure}[h]
\setlength{\unitlength}{1.2mm}
\begin{center}
\begin{picture}(60,10)
\put(0,4){\vector(1,0){60}}
\put(15,2.5){\line(0,1){3}}
\put(30,2.5){\line(0,1){3}}
\put(45,2.5){\line(0,1){3}}
\put(15,0){\mbox{$t_e$}}
\put(30,0){\mbox{$t_c$}}
\put(40,0){\mbox{$t_H\!=\!\frac1{\Delta}$}}
\put(59,0){\mbox{$t$}}
\put(2,7){ballistic}
\put(17,7){diffusive}
\put(33,7){ergodic}
\put(47,7){quantum}
\end{picture}
\end{center}
\caption{Schematic ordering of the diffusion time-scales.}
\end{figure}

In metallic grains, in the ergodic window the spectral properties of
the
metals are universal, and described by the random matrix theory
(universal conductance fluctuations), hence the second name of this
regime. 

Let us try to make this description more quantitative. Let us
introduce the retarded
Green's function $G^R (x,y;E)=<x|(E-H +i\epsilon)^{-1}|y>$.
The time-Fourier transformation $ G^R(x,y; t)$ describes the
amplitude, that the diffusing particle propagates from $x$ to $y$ 
in a time $t$, under the influence of the dynamics governed by some Hamiltonian $H$.
The Hamiltonian is the microscopic source of the disorder.
A classic example is the Anderson Hamiltonian~\cite{ANDERSON}
\be
H_{A}=-\frac{1}{2m}(\nabla-ieA)^2 +V(\vec{x})
 \ee
where $<V(\vec{x})>=0$ and
$<V(\vec{x})V(\vec{y}>=\delta(\vec{x}-\vec{y})/(2\pi \bar{\rho}(E_F)t_e)$.
Note that the disorder is {\em static}, i.e. time independent.
We can now define a crucial concept of the return probability.
This is simply the square of the amplitude of returning to the same
point $x$ in a time $t$, averaged over the disorder.
For a particle at a Fermi surface, it reads
\be
P(t) =\lim_{x\rightarrow y}\frac{V}{2\pi\bar{\rho}} \int d\lambda
e^{i\lambda t}\left<G^R(x,y;
  E_F+\lambda/2) [G^R(x,y;E_F+\lambda/2)]^*\right>_{H_A}
 \label{retmet}
\ee
The prefactor $V$ comes from the translational  invariance of the
return probability, and the denominator $2\pi \bar{\rho}(E_F)$ guarantees
the normalization of the probability to 1. 

For a {\em static} random potential of the type above one can perform the
averaging and then the  integration, with the following result:
\be 
P(x,y,t)=\sum_{\vec{q}} e^{D q^2 t} e^{i\vec{q}(\vec{x}-\vec{y})}
\stackrel{\lim_{V \rightarrow \infty}}{\longrightarrow} 
\frac{V}{(4\pi D t)^{d/2}} e^{-|\vec{x}-\vec{y}|^2/4Dt}
\ee
Hence
\be
P(t)=\sum_{\vec{q}} e^{-Dq^2t}
 \ee
or, equivalently, in Fourier space 
\be
P(\lambda)=\sum_{\vec{q}} \frac{1}{-i\lambda +D q^2}
\ee 
In the above, while performing the averaging over the microscopic disorder, 
we integrated out fast degrees of freedom, getting the effective
description in terms soft modes $q_i$. Details of fast degrees of freedom are
now hidden in effective parameters like the diffusion constant $D$, 
and the obvious name "diffuson" for a soft modes $q_i$ is natural in the
light of the form $P(x,y,t)$, being the Green's function of a diffusion
operator $(\partial_t + D\nabla^2)$.
 
Formally, for very large times ($\lambda \rightarrow 0$), the return
probability
develops a pole (diffuson pole at $q^2=0$). However, for such times,
quantum effects become relevant,
 introducing the natural cutoff $\gamma$ at the 
energy scale of the average quantum spacing $\Delta$. This cutoff suppresses 
infinitely long diffusive orbits, and regulates the pole
\be
P(t)=e^{-\gamma t}\sum_{\vec{q}} e^{-Dq^2 t}
\ee
or, after Fourier transformation
\be
P(t)=\sum_{\vec{q}} \frac{1}{-i\lambda +\gamma +Dq^2}
\ee
In the next section, we will demonstrate, that all the above listed
concepts
are directly applicable to  QCD.

\subsection*{ Euclidean QCD is diffusive in d=4}

Let us look at the Dirac operator in the background of some 
Euclidean gluonic configuration $A$
\be
(iD\!\!\!\!/(A) +im) q_k =\lambda_k(A) q_k
\ee
as  a "Hamiltonian" corresponding to the "eigenenergy"  $\lambda$. 
The imaginary constant shift $im$ does not spoil the analogy.
For static Hamiltonians in $d=1,2,3$ we know from the quantum mechanics, 
that the evolution in time $t$ is governed by $q_k(t)=\exp (iE_kt)
q_k(0)$.
Time $t$ is dual to the energy $E$.
We will parallel this construction here, introducing dual "time" $\tau$
to the virtuality ("energy") $\lambda$. Since the "Hamiltonian"
$iD\!\!\!\!/(A)+im$
is by definition independent on this "time" ({\em static} in $\tau$), we can
write down the eigenmode evolution in this "time"
\be
q_k(\tau) =e^{ i (\lambda_k+im)\tau} q_k(0)
\ee
corresponding to $4+1$ dimensional Schroedinger-like  equation
with static ($\tau$ independent) potential
\be
i\partial_{\tau}q(\tau)= (iD\!\!\!\!/(A) +im )q(\tau). 
\ee
Note that this construction does not modify any properties of the
Dirac
operator, and is basically equivalent to introducing the  Schwinger
proper time $\tau$. 

We will prove now, following~\cite{PRLUS},
that the dynamics of the Euclidean $d=4$ Dirac operator in
Schwinger time $\tau$ parallels the dynamics of the usual diffusion $d=1,2,3$ 
in the real time $t$. This means, that we will identify, in the
spectrum 
of the Dirac operator, four distinct regimes corresponding
to quantum, ergodic, diffusive and ballistic regimes of the mesoscopic
physics.

From now on, we follow,  step by step, the construction outlined in the
previous chapter. Instead of Green's function for a diffusing electron  
we take a Green's function for a  quark in some unknown
gluonic background, which is constant  in Schwinger time $\tau$
\be 
G^R(x,y;E) \longrightarrow   S_F (x,y;\lambda)
\ee
We can now define by analogy
  to (\ref{retmet}) the return probability in time $\tau$
for a quark at the {\em zero virtuality surface} 
\be
P(\tau)=\frac{V}{2\pi \bar{\rho}(0)}\int d\lambda e^{i\lambda \tau}
 \left< {\rm
    Tr}\,\, S_F S_F^{\dagger}\right>_{\rm QCD}
\ee
Here $S_F\equiv S_F(x,y;m-i\lambda/2)$, trace appear due to the $\gamma_{\mu}$
matrix structure of the Dirac operator and the averaging is done over
the full QCD measure (\ref{QCDweight}), representing the analog of
Anderson Hamiltonian
representing the disorder of the system. The
 parameter $m$ appears here  due to the uniform  shift $im$ of
the original spectrum of the massless Dirac operator. 

Till now the analogy was exact. But now we have to perform the
averaging over the a priori unknown measure of the QCD.
The first alternative is to choose some model of the disorder.
Actually the analytical instanton model of Diakonov and Petrov~\cite{DIAKONOVPETROV}   
 and the numerical simulation by Shuryak~\cite{SHURYAK} are the first
 realizations of
 this scenario. We will return to this point later. 

Surprisingly,  we can  calculate $P(\tau)$ {\em without assuming any model
of the disorder}, but making use of almost exact low energy theorems 
of the QCD, introduced in Part~1.
First, let us note, that due to  the chiral properties of the Dirac 
operator
\be
<{\rm Tr}S_FS_F^{\dagger} >= -<{\rm Tr} S_F(x,y;z)\gamma_5 S_F(y,x; z^{*}\gamma_5> 
\ee
where $z=m-\lambda/2$. We may introduce also  the isospin
sources.  One then immediately recognizes the similarity to the pion
correlation function structure
\be
 C_{\pi}^{ab}(x,y;m) &=& <S_F(x,y;m) i\gamma_5 \tau^a S_F(y,x;m)
 i\gamma_5 \tau^b>_{\rm QCD} \nonumber \\ &=&
\frac{1}{V}\sum_{Q_{\mu}} e^{iQ(x-y)}\left[ \frac{<\bar{q}q>}{F_\pi}\right]^2
\cdot \frac{1}{Q^2+M_{\pi}^2}
\ee
where we used the pion dominance formulae (\ref{pseudoscalar}). 

Second, we use another low-energy theorem (\ref{GOR}), to replace the
$M_{\pi}$ on the r.h.s. of the above equation by the current quark
mass $m$. 
Third, we analytically continue the $C_{\pi}^{ab}(x,y;m)$, by replacing
the mass $m$ by $z=m-i\lambda/2$. We recognize then  
that the integrand  in the return probability
involves the analytically continued pion correlation function, so
averaging is 
equivalent to the analytical continuation of the r.h.s. of $C_{\pi}$
\be
 P(x,y, \tau)&=&\frac{1}{2\pi \bar{\rho}}\int d\lambda e^{i\lambda \tau}
C_{\pi}(x,y;
 z(\lambda)) \nonumber \\
 &\approx& \sum_{Q_{\mu}} \int d\lambda e^{i\lambda \tau} 
e^{iQ(x-y)}  \frac{1}{\frac{F^2_{\pi}}{|<\bar{q}q>|} Q^2 + 2m +i \lambda}
\ee
 
\noindent Finally, after integrating  over $\lambda$ by the residue method,
we see already the diffusive structure of  
$P(x,y,\tau)$,
and  the return probability (limit $x \rightarrow y$) reads  
\be
P(\tau)=e^{-2m\tau} \sum_{Q_{\mu}} e^{-DQ^2\tau}
\ee
or, after Fourier transforming
\be
 P(\lambda)=\sum_{Q_{\mu}} \frac{1}{-i\lambda +2m +DQ^2}
\ee
We constructed in this way 
 the precise analog of the diffusive return
probability, with diffusion constant $D=F^2_{\pi}/|<\bar{q}q>|\sim
0.22$ fm  and the slow "diffuson
modes"
$Q_{\mu}=2\pi n_{\mu}/L$, with $\mu=1,2,3,4$ and $n_{\mu}$ integers.

In fact, using the low energy theorems and the above construction,
we {\em implicitly} integrated out the fast (gluonic) degrees of freedom, 
getting in this way the diffusion constant. 
The remaining dynamics of slow modes (diffusons) is the dynamics
of the pions - longest wave excitations of the QCD.
The Goldstone nature of the pion manifest itself as a pole in the
chiral limit and $\lambda \rightarrow 0$ limit. 
We see  easily that the very heavy quarks do not diffuse at all
(due to the exponential damping).  Note also that for light but
massive quarks we can read  GOR as 
an expression of the coherence length of the QCD vacuum. 
Indeed, by definition, coherence length is 
\be
L_{coh}\equiv \sqrt{D/\gamma}=1/M_{\pi} 
\ee
where we used that for QCD $\gamma=2m$ and 
the expression for the diffusion constant  calculated
above. Coherence length of the QCD vacuum is related to the pion mass,
so 
in the massless limit, pions are indeed much more vacuum modes
than the pairs of bounded constituent quark and antiquark.

Now we can identify the different regimes in the spectrum of the Dirac
operator. Since the Schwinger time $\tau$ is dual to the eigenvalue
of the Dirac operator, we immediately identify the analog of Thouless
energy, here the Thouless virtuality~\cite{PRLUS,VERBOS}
\be
\Lambda_c=D/L^2=F_{\pi}^2/(\Sigma \sqrt{V})
\ee
where $|<\bar{q}q>| \equiv \Sigma$. 
Therefore the eigenvalues smaller than $\Lambda_c$ are expected to
belong to the ergodic (universal) regime, whereas larger than $\Lambda_c$ to the
diffusive. Since the Banks Casher relation gives us an estimate of the 
mean level spacing, we know also the borderline of the quantum and
ergodic regime. Finally, for very short times ( large eigenvalues)
the concept of the diffusion becomes meaningless. It is not very
difficult to argue, that the borderline between the diffusive and
ballistic regime is given by twice the mass of the constituent quark,
basically the mass of the $\rho$ meson.
Indeed, for times shorter than the required to travel one mean-free
path between lumps of the gauge field  the concept of  dressing ( via  multiple scatterings) 
 of the current mass is void. 

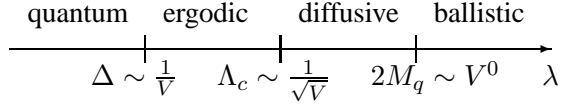
\begin{figure}[h]
\begin{center}
\setlength{\unitlength}{1.2mm}
\begin{picture}(60,10)
\put(0,4){\vector(1,0){60}}
\put(15,2.5){\line(0,1){3}}
\put(30,2.5){\line(0,1){3}}
\put(45,2.5){\line(0,1){3}}
\put(9,0){\mbox{$\Delta \sim \frac{1}{V}$}}
\put(23,0){\mbox{$\Lambda_c \sim \frac{1}{\sqrt{V}}$}}
\put(40,0){\mbox{$2M_q \sim V^0 $ }}
\put(59,0){\mbox{$\lambda$}}
\put(2,7){quantum}
\put(17,7){ergodic}
\put(32,7){diffusive}
\put(47,7){ballistic}
\end{picture}
\end{center}
\caption{Disorder regimes in the eigenvalue 
spectrum of the Dirac operator. Note the thermodynamical ordering. }
\end{figure}

The above hierarchy of scales could be summarized again by the cartoon, 
where we also remind  about the thermodynamical ordering of the
scales. 
On the basis of the analogy with the condensed matter systems, 
we may suspect, that similarly to the universal conductance
fluctuation in the ergodic regime, we may see universal spectral
fluctuations in the ergodic window of the QCD, hopefully described by
some
sort of random matrix theory. 
In the next chapters, we will see that this indeed is the case.  
 
\subsection*{ Ergodic ( universal)  regime of the QCD}
Let us consider in more detail the ergodic regime. 
We expand the sum over the diffuson (pion) modes valid in diffusive
regime, i.e. we sum over the quadruples of integers $Q_{\mu}=(n_1,n_2,n_3,n_4)2\pi/L$
\be
P(\tau)=e^{-2m\tau}(1 + 8e^{-4\pi^2 D/L^2 \tau } +...)
\ee
where the first term comes from all $n_i=0$, second from all
combinations of $n$ of the type $(1,0,0,0)$ etc. 
Since $D/L^2=\tau_c^{-1}$, the return probability reads 
\be
P(\tau) = e^{-2m\tau} (1 + 8e^{-4\pi^2 \tau/\tau_c} + ....)
\ee
For the times larger than $t_c$ all terms except of the first one
vanish  exponentially, so we obtain, that in the ergodic regime 
the return probability is simply
\be
P(\tau)=e^{-2m\tau}  
\ee
approaching constant for very large $\tau$.
Only the softest modes (zero modes $n_1=n_2=n_3=n_4=0$) determine the
return probability in the ergodic regime.  This points that the
properties  of the ergodic regime are universal -- they are independent 
on the details of space-time interactions, since the Goldstone 
bosons interaction involves derivative terms, which vanish in long
range limit. 

This result was known since long ago, although rephrased in a  
different way. In the usual chiral perturbation theory,
pion  momenta are of order $1/L$, the mass of the pion scales
therefore
like $1/L^2$, and since $m\sim M_{\pi}^2$, the combination 
$m^2V \sim O(1)$ is fixed.  The systematic expansion based on this
counting is the clue of the Weinberg~\cite{WEINBERG} chiral perturbation expansion
and was practically realized  by Gasser and
Leutwyler~\cite{GASSERLEUTWYLER}. 
What is less known, that Gasser and Leutwyler looked also at finite
volume, 
and  what 
happens, when one keeps only zero modes of the pion propagator.
Since the propagator in the final volume reads
\be
G&=&\frac{1}{V}\sum e^{iQ_n x} \frac{1}{Q_n^2 +M_{\pi}^2} \nonumber
\\
&=&
 \frac{1}{VM_{\pi}^2} + \frac{1}{V} \sum {}^{'}
e^{iQ_n x}\frac{1}{Q_n^2 +M_{\pi}^2}
\ee
the counting based on the concept of having $mV$ fixed~\cite{MOREL}(equivalent to
$ M_{\pi}^2V \sim O(1)$) 
kills all the terms in the primed sum (non-zero modes) and leaves the zero mode. 
This is very different  to the counting based on before-mentioned $m^2V=1$
principle (chiral perturbation theory). 

Gasser and Leutwyler
 managed to re-sum the contribution of all the
zero modes based on  $mV\sim O(1)$ counting, and obtained the exact partition
function
\be
Z(m)=\int dU e^{V\Sigma {\rm Tr}\, m(U +U^{\dagger})}
\ee
with $U=\exp (i\pi\tau)$ and $\Sigma=|<\bar{q}q>|$. 
 This formula is universal: it depends only on the ways how chiral
symmetry is to be  broken spontaneously (choice of the measure $dU$ of
the Goldstone modes)   
  and how chiral symmetry is being broken explicitly, by 
the $(N_f,\bar{N}_f) + (\bar{N}_f,N_f)$ representation in the 
exponent.  

In the ergodic regime any theory/model  sharing the same global
symmetries as QCD belongs to the same universality class, 
i.e. any theory/model leading to the return probability 
$P(\tau)= e^{-2m\tau}$  will exhibit the same universal spectral
properties in the ergodic window.
Since the information about the dimensionality is being lost in the 
ergodic regime, this could be even a zero-dimensional version of the 
field theory.  But field-theory in zero-dimensions is a
matrix model. So in the ergodic window QCD is equivalent to a  certain
matrix model. Before we will unravel the details of this model,
let us remind once more that this equivalence between QCD and Random
Matrix Model (RMM) happened
only for eigenvalues  {\it below}  $\Lambda_c$. 
Since the edges of the ergodic  window scale with volume  as $\Delta \sim 1/V$
( Banks-Casher) and as $\Lambda_c \sim 1/\sqrt{V}$ ( Thouless
virtuality in four dimensions)
 the ergodic, universal window shrinks to a point 
 in the infinite volume limit.

\subsection*{ Where does the "color dirt" come from?}

At this moment we may start to worry, what plays the role of the 
diffusive dirt
in the case of the QCD? 
QCD is a fundamental theory, so the "dirt" has to be an immanent
feature of the QCD itself. Whereas we do not have an exact answer what is
the color dirt in QCD, let us observe, that most of the 
localized, Euclidean (i.e. static from the point of view of Schwinger
time) gluonic configurations do the job. 
Typical and perhaps the most natural are instantons~\cite{INSTANTONS}. 
First, they are the classical, localized 
 stable Euclidean solutions of the QCD. 
Second, Dirac equations in the presence of instanton background 
possesses  the chiral zero modes, therefore the instanton vacuum immediately
provides a microscopic "hopping" mechanism from one instanton field
to another and guarantees flipping  the chirality at each 
"scattering" in "time" 
$\tau$. Quantitatively, each instanton provides a seed 
of non-conservation of the chiral charge, 
by integrated form of the anomaly:
\be
\delta Q_A =2N_f Q
\ee
where $Q$ is the topological charge of the instanton. 
The instanton model involves basically two parameters, the average
size of the instanton ($\rho \sim 1/3$ fm) and the concentration
$n\sim fm^{-4}$
(the typical density of instantons). 
 Therefore every dimension-full quantity in the instanton model depends
 parametrically on the combination of the two scales. 
In particular, the smallness of the diffusion constant calculated in
the previous chapter, finds a natural explanation as a diluteness of
the instanton medium. Note that the finite value of the condensation 
requires that thermodynamical limit $V\rightarrow \infty$,
 $N \rightarrow \infty$, with $N/V$ fixed,  where  $N$ is the number of
instantons
and antiinstantons.

Each instanton vacuum configuration is a snapshot in a time $\tau$.
Each snapshot violates the Lorentz invariance ("static", 
particular  distribution 
of instantons and antiinstantons in four volume)  and gauge invariance
(each instanton freezes a direction in color). 
But the averaging over collective coordinates of the instantons (here
over the centers of instantons and over their color measure)
restores  the gauge and Lorentz invariance.
 We would like to stress, that instantons are sufficient, but not
 necessary
configurations to realize the diffusive scenario. 
Several other models may also provide  chiral disorder,
e.g. family of stochastic vacuum models~\cite{HEIDELBERG}. 
Each of these  models comes with a certain correlation length scale, 
which corresponds quantitatively  to  the diffusion constant.

In other ways, the spontaneous breakdown of the chiral symmetry 
in QCD is a very robust phenomenon, comparing to the confinement.
Usually, the models of color dirt either ignore confinement, or
introduce it by hand, or as instantons, seem to be (at least naively) 
not related to confinement at all. 
  Mysteriously, lattice evidence suggests
strongly that both phenomena ( confinement and chiral symmetry
breakdown)
are  correlated and vanish at the same temperature.
Since the confining configurations are the topic of vivid speculations, 
the fundamental  understanding of the "color dirt" and chiral disorder
is still missing.

\section*{Part  3}
I  analyze in more detail two regimes of the
QCD, 
the ergodic one  and the diffusive one. Then I  confront  few
sample 
predictions
with the "experimental data" obtained from lattice simulations.    

\subsection* {Random matrices - field theory in 0 dimension}
As stated before, in the ergodic regime any model obeying the global
symmetries of the QCD belongs to same universality class as QCD. 
Since this is the regime where only the constant pionic  modes matter,
we can ignore space-time dependence and stick to the field theory 
in zero dimensions, i.e. the theory where fields are numbers and do
not have any space-time dependence. 
The exact form of such theory in the QCD can be inferred in  numerous
ways.
Here we follow the historical
route~\cite{DIAKONOVPETROV,SHURYAK,OLDUS}.
 Imagine that we have two lumps  
of the "dirt", e.g. instanton and anti-instanton, separated at the
very large distance. Then, the Dirac equation in the field of the
instanton  has an exact fermionic right-handed zero mode $\phi_R$, and the 
Dirac equation for the anti-instanton has a similar, but left-handed
zero mode $\phi_L$. When we decrease  the distance between instantons, the
degenerated pair of zero modes is replaced by the pair of eigenvalues
($ T, -T$), where the overlap $T=\int d^4x \phi_R^{\dagger} i\partial
\!\!\!/ \phi_L$ depends on the distance and the mutual orientation 
of the instantons. Let us now add more and more instantons into 
this medium, and still work in a dilute gas approximation. 
The  infinite fermion determinant, when calculated in the basis 
of left and right handed quark zero modes,  is now approximated 
by the matrix of overlaps between the I-th instanton and J-th
anti-instanton $T_{IJ}$. The off-diagonal block structure comes 
from the chirality flipping mechanism. The diagonal blocks are  zero in
the chiral limit. In this way, instanton picture trades  the a priori
unknown QCD measure (\ref{QCDweight}) into the approximate measure
\be
<(...)>=\int \prod d[\Omega_i](\ldots) e^{- S_{\rm glue}(\Omega_i)} 
\prod_{N_f} {\rm det} 
\left| \begin{array}{cc}
0 & T(\Omega_i)\\
T^{\dagger}(\Omega_i) &0 \\
\end{array} \right|
\ee
where $\Omega_i$ is the set of collective coordinates of the
instantons ( positions and color orientations), and $S_{\rm glue}$
is the gluonic part of the QCD action saturated with the initial
instanton vacuum ansatz. Let us truncate now all the space time
dependence
in the above action.
We are left with the model of the type
\be
Z=\int dT e^{- N\sigma {\rm Tr} T T^{\dagger}} \det \left| \begin{array}{cc}
0 & T\\
T^{\dagger} &0 \\
\end{array} \right|^{N_f}
\label{zchir}
\ee
Here $T$ is an $N$ by $N$ matrix built out of complex numbers (one
could generalize the matrices to the rectangular ones as well). 
The gluonic measure is replaced by the 
polynomial measure with some potential $v(T^{\dagger}T)$, 
here being the simplest -
the 
harmonic potential.
The factor $N$ in front of the potential guarantees, that
each integration over $T$ is appropriately weighted. The model has one
scale. i.e. the "width" $\sigma$ of the Gaussian matrix measure.
The partition function $Z$ defines  the chiral Gaussian random matrix 
model~\cite{SHURVER}.   
The name chiral comes from the off-diagonal block structure,
which mimics the original chiral structure of the Dirac operator, 
$[D\!\!\!\!/(A),\gamma_5]_+=0$.

We will show now, that such defined model is exactly solvable 
for any finite $N$. 
The main problem is the integration over the measure $dT$. 
We will use here the trick, similar to the change of the variables 
from Cartesian coordinates to the spherical ones. 
The simplest way to change usual coordinate 
variables is to look at the infinitesimal
interval
\be 
ds^2= dx^2 +dy^2+dz^2= dr^2 +r^2 d\theta^2 +r^2 \sin^2\theta d\phi^2
\ee
so the metric tensor is $g_{lk}={\rm diag} (1, r^2, r^2 \sin^2\theta)$.
Hence the Jacobian (the square root of the determinant of the metric
tensor) reads
  $J=\sqrt{\det g_{lk}}\equiv \sqrt{g}=r^2 \sin \theta$. 
We observe now that the integrand of the partition function (\ref{zchir}) 
depends only on the
"radial"
combination $X=T^{\dagger}T$, and we follow  the 
trivial example above. 
Every hermitian matrix   can be diagonalized by a  unitary
transformation. Since $X$ is hermitian, introducing $X=URU^{\dagger}$, where 
the unitary matrix $U=\exp (i H)$, $H$ hermitian, 
 and $R_{ij}=\delta_{ij}r_i $ is positive
diagonal, 
we calculate
\be
{\rm tr} (dX^2)&=&{\rm tr} (U^{\dagger} (dR+i[R,dH])U)^2 =
     {\rm tr} (dR +i[R,dH])^2 
 \nonumber \\       &=&   \sum_k dr^2_k
+\sum_{i \neq j}(r_i-r_j)^2 |dH_{ij}|^2
\ee
Hence the metric tensor reads
\be
g_{lk}=  {\rm diag}( \underbrace{1,......1}_{N},\underbrace{..., (r_i -r_j)^2, ....}_{N(N-1)})
\ee
so det$ g_{lk} =\prod_{i \neq j}(r_i-r_j)^2$ and Jacobian 
$J=\sqrt{g}=\prod_{i <j} (r_i-r_j)^2\equiv  \Delta(R)^2$. 
Since the integrand and the Jacobian 
do not depend on the angles parameterizing the 
unitary matrices, we are left with $N$ integrations over the diagonal 
eigenvalues  ("radial" variables) $r_i$
\be
Z=\int \prod_i dr_i \Delta(R)^2 \exp (-N \sigma \sum r_i)\prod  r_i^{N_f}
\ee
Let us note, that the expression $\Delta(R)$ can be rewritten as a
determinant
(Vandermonde  determinant).
\be
\Delta(R)= \prod_{i < j} (r_i-r_j)=\left| \begin{array}{ccccc}
1 & r_1 & r_1^2 & \ldots & r_1^{N-1} \\
1 & r_2 & r_2^2 & \ldots & r_2^{N-1} \\
\vdots & \vdots & \vdots & \ddots & \vdots \\
1 & r_N & r_N^2 & \ldots & r_N^{N-1}\\
\end{array} \right|
\label{Vdm1}
\ee
[\underline{Exercise}. {\it Prove }(\ref{Vdm1}).] 
Since we can add to each row of the determinant  an arbitrary combination
of the other rows (without changing the value of the determinant), 
we can replace the original Vandermondian by the determinant build out
of polynomials. 
\be
\Delta(R)=\left| \begin{array}{ccccc}
P_0(r_1) & P_1(r_1) & P_2(r_1) & \ldots & P_{N-1}(r_1) \\
P_0(r_2) & P_1(r_2) & P_2( r_2) & \ldots & P_{N-1}(r_2) \\
\vdots & \vdots & \vdots & \ddots & \vdots \\
P_0(r_N) & P_1(r_N)& P_2(r_N) & \ldots & P_{N-1}(r_N)\\
\end{array} \right|
\ee
These polynomials are a priori arbitrary, modulo the
fact
that the coefficient at the highest power is always 1. 
 We choose now the polynomials in such a way, 
that they are orthogonal with respect to the measure. In our case we
require, 
that 
\be 
\int_o^{\infty}  dr e^{-N\sigma r} r^{N_f} P_i(r)P_k(r)=\delta_{ik}
\ee
It is obvious, that in the case of $N_f=0$ such polynomials are
Laguerre
polynomials (modulo trivial rescaling of the coefficients, so the 
highest power is multiplied by 1)  and in case of $N_f \neq 0$ 
are associated Laguerre
polynomials $L^{(N_f)}_k$. Due to the orthogonality of the polynomials all
integrations in the partition function factorize, and we can write
down
arbitrary $k$-correlation function  between the $k$ eigenvalues.  Integrating
over all eigenvalues except of the last one,
we arrive at the average spectral density
\be
\rho(\lambda)&=&\sum_{k=0}^{N-1} 
\frac{k!}{\Gamma(N_f+k+1)} L_k^{(N_f)} L_k^{(N_f)}
  z^{N_f+1/2}
e^{-z} 
 \nonumber \\
&=& \frac{N! z^{N_f+\frac{1}{2}}e^{-z}}{\Gamma(N_f +N)}
 \left[L_{N-1}^{(N_f)}L_{N-1}^{(N_f+1)}-
L_N^{(N_f)} L_{N-2}^{(N_f+1)} \right]
\label{dens}
\ee
where we used (in the second line) the Christoffel-Darboux summation
formula and where $z=N\lambda^2\sigma$ is an argument of the
polynomials,
with $\lambda$ being the original 
eigenvalue of the matrix $T$. 

Let us now come back to physics. In the previous section we stressed, 
 that the ergodic regime corresponds 
to the limit, where $Vm$ is fixed (in units of condensate $\Sigma$).  
This suggests, that the relevant
limit for ergodic QCD follows from the chiral random matrix model with 
$N\lambda$  fixed (in units of the width $\sigma$)~\footnote{The rigorous proof of the equivalence 
of RMM limit and chiral partition function is
 non-trivial~\cite{DAMGAARDCHI}.}.
 Technically this is achieved by so called unfolding
procedure. We  define the microscopic spectral density
$\rho_{\rm micro} (s) \equiv \lim_{N \rightarrow \infty} \frac{1}{N\sigma}
\rho(\lambda \equiv \frac{s}{N\sigma})
 $.
Using the asymptotic property of Laguerre polynomials 
\be
\lim_{N \rightarrow \infty} \frac{1}{N^{a}}  L_N^{(a)} (x/N)=
x^{-a/2} J_a (2\sqrt{x})
\ee
we rewrite (\ref{dens}) in the microscopic limit in terms of the
following combination of the Bessel functions
\be
\rho_{\rm micro}(s)=\frac{1}{2}\sigma^2s (J^2_{N_f}(\sigma s)
-J_{N_f-1}(\sigma s) J_{N_f+1}(\sigma s))
\label{micro}
\ee
This formula~\cite{VERBAARSCHOTZAHED}
 represents the universal fluctuations ( universality
class
of chiral Gaussian unitary ensemble) of the 
spectral density in the QCD. The parameter $\sigma$ is fixed  by 
tuning the dimensionless quantity $N\lambda \sigma$ to 
dimensionless quantity $Vm\Sigma$.  
It is important to mention, that this formula is universal,
i.e. similar
for any polynomial potential~\cite{DNM} $v(T^{\dagger}T)$ of the random matrix model.   
If  we would not impose the limit  $N\lambda=O(1)$
while $N\rightarrow \infty$, instead of $\rho_{\rm micro}(s)$ given by 
(\ref{micro}) we would  have obtained the non-universal 
spectral density of the Gaussian random models, i.e. Wigner
semicircle $\rho(\lambda) \sim \sqrt{ 4-\lambda^2}$. 
\begin{figure}[htbp]
\centerline{\epsfysize=55mm \epsfbox{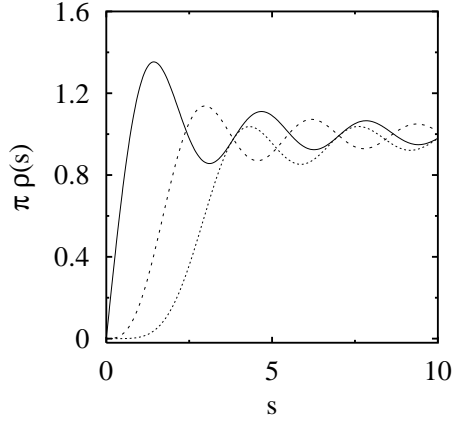}}
\caption{Microscopic spectral density for $N_f=0,1,2$ (solid, dashed
  dotted lines, respectively).}
\end{figure}
Colloquially speaking, all the universal fluctuations are hidden at 
a single point of the Wigner semicircle, 
 corresponding to point   with coordinates $(\lambda=0, \rho(0))$. 
The "zooming"  of this micro-dot by factor $N$ unravels the fluctuating,
universal
pattern of microscopic spectral density of the chiral ensemble (Fig.~3).

\subsection* { Crossover to the diffusive regime - chiral perturbation} 
The universal (ergodic) regime  ends at the scale corresponding to
the Thouless
virtuality. For the scales larger than Thouless virtuality 
neglecting  non-zero modes is no longer justified -- 
all modes have to be taken into account.
Let us look  at the scaling properties of the eigenvalues.
The first eigenvalue  which falls off from the ergodic window
(i.e. by definition,  Thouless virtuality) 
has to scale~\cite{PRLUS,VERBOS} (in units of 
$\Delta$) as 
\be
\frac{\Lambda_c}{\Delta}= \frac{D}{\Delta L^2}=
\frac{F^2_{\pi}}{\Delta |<\bar{q}q>| L^2} \sim  \frac{F_{\pi}^2 V}{L^2} =
  F_{\pi}^2 L^2 
\label{conductance}
\ee
where we used the Banks-Casher relation and the definitions of the
Thouless
virtuality and the diffusion constant.

It is remarkable, that the analogy with the diffusion can lead to a
quantitative description of the spectral correlations in the
 diffusive 
 regime of the QCD as well.
Let us define two-point correlation density of states
\be
R(s)=\frac{1}{<\rho(\Lambda)>^2}\left<
  \rho(\Lambda-\epsilon/2)\rho(\Lambda+\epsilon/2)\right>-1
\label{dosdos}
\ee 
where $s=\epsilon/\Delta$ and, as usual, $\Delta \sim 1/\rho V$.
Then we define the spectral formfactor
\be
K(t)=\frac{1}{2\pi} \int d\epsilon R(\epsilon)e^{-i\epsilon t}
\label{spectral}
\ee
For times shorter than the Heisenberg time $t_H\sim \Delta^{-1}$, the 
standard  semi-classical argument  developed for mesoscopic systems 
by~\cite{ARGAMAN}
allows 
to relate spectral formfactor to the return probability $P(t)$
\be
K(t)= \frac{1}{4\pi^2}t \Delta^2 P(t)
\label{Berry}
\ee
By definition, the two-level 
spectral correlation function, integrated  from $0$ to $\Lambda$ 
measures the fluctuation $\Sigma^2(\Lambda)$ of the number of levels 
$N(\Lambda)$ 
in a strip of a width $\Lambda$
\be
\Sigma^2(\Lambda) =\int_0^{<N>} \int_0^{<N>} 
d\lambda_1 d\lambda_2 R(\lambda_1-\lambda_2)
\ee 
In the ergodic regime (for $t \rightarrow \infty$)
 $P(t)$ is a constant, hence $K(t) \sim t$.
The variance $\Sigma^2$ grows 
as logarithm of $\Lambda$, as expected in the random matrix theory.  
  However, for times smaller than
$t_d$ ( eigenvalues greater then Thouless energy), 
the return probability is given by the classical diffusion result
\be
P(t)= \frac{V}{(4\pi Dt)^{d/2}}
\ee
Then $K(t) \sim t^{1-d/2}$, which corresponds, 
after Fourier transforming, to 
\be
R(s) \sim 
 - {\rm Re\,\,\,}{\Delta^2} \sum_{Q} \frac{1}{(s\Delta +iDQ^2 )^2}
\label{alts}
\ee
For  mesoscopic systems (e.g. for 
disordered electrons in $d=3$ metallic grains), 
this is the seminal  result obtained 
by Altshuler and Shklovskii~\cite{ALTSCHULER}, 
demonstrating, that the two-point
correlation function comes diagrammatically from the two-diffuson
exchange. The variance $\Sigma^2(E)$ reads then 
\be
\Sigma^2(E)\sim \left(\frac{E}{E_c}\right)^{d/2} \sim \left(\frac{L}{L(E)} \right)^d
\ee
The second equation comes from definition of the Thouless scale and
from
$ E \sim t^{-1}=D/L^2(E)$. 
The {\it power behavior} reflects the fact, that in the diffusive
regime  diffusion of a particle with energy $E$ is non-homogeneous, 
and  
takes place independently in the number of 
sub-blocks $(L/L(E))^d$ of the original block $V=L^d$.

The above mentioned crossover from universal regime ( logarithmic behavior of 
the variance) to diffusive regime ( power behavior $(E/E_c)^{3/2})$
was  confirmed by~\cite{BROWNMON} in numerical simulation
of  the metallic regime for Anderson model with $20^3$ sites.

On the basis of the diffusive scenario presented in these notes, 
we expect similar behavior for QCD, but with $d=4$ and $t \rightarrow \tau$. 
In our case, the square denominator in (\ref{alts}) follows from the
exchange of two pions in the double ring diagram corresponding to
density-density correlation function, reflecting on  disconnected quark
susceptibility. Note that we have obtained this result on the basic of
spectral analysis, 
and not 
by using a  standard diagrammatics
 of the  chiral perturbation.   
The diffusive regime of  Altschuler-Shklovskii in  mesoscopic systems is
equivalent to chiral perturbation theory. 
The effective models of diffusons (sigma models) correspond
to sigma models of the pions, and supersymmetric formulation  of the
diffuson models by Efetov~\cite{EFETOV} resembles 
  the family of so-called (partially)
quenched chiral perturbation theories~\cite{MARTIN}.  
Finite volume QCD plays a role of a 4-dimensional {\it quantum dot}.

For QCD, the diffusive regime should  show the 
power behavior $\Sigma^2(\lambda)  \sim (\lambda/\Lambda_c)^2$, since the 
diffusion takes place in $d=4$ (~\cite{PRLUS}).
Recent  lattice simulation~\cite{TILOLAST} of the $SU(3)$ QCD has confirmed 
this scenario, demonstrating clearly the expected crossover from
ergodic regime
(logarithmic behavior) to diffusive regime (power-like behavior)
predicted for the QCD.
It is illuminating  to compare the character of the crossover  
in metallic grains (e.g.  Fig.~3 in~\cite{BROWNMON})
 to crossover regime of the
 disconnected  quark susceptibility (e.g. Fig.~8  in~\cite{TILOLAST}). 

Let me  mention another intriguing analogy. 
In the case of Hamiltonians invariant under the time reflection, one does
not differentiate between the direction, in which the loop contributing
to return probability is traversed. 
Hence the naive contribution to the return probability is doubled, 
which corresponds to 
additional diffuson-like contribution in the formula for return probability. 
This quantum interference of the identical orbits 
traversed in opposite directions  is called in condensed matter
 {\it coherent backscattering
(weak localization)}, 
and the collective diffuson like-excitation corresponding to this
effect is called a {\it cooperon}, since the charges add during this
interference.
This phenomenon has also  an analogy in QCD~\cite{PRLUS,USEXTERNAL}.
 In case of a  real color
group
(e.g. $SU(2)$), we do not distinguish between the quark and antiquark,
and on top of the usual pions (diffusons) we can form the collective
states corresponding to $qq$ pairs, so called baryonic pions - 
hence QCD-cooperons. The above mentioned lattice 
analysis~\cite{TILOLAST} clearly identified
the baryonic pions in quenched lattice simulation, in agreement with 
theoretical predictions based on the diffusive scenario of the QCD.  

In the light of the above results, it is tempting to speculate, 
that the ideas borrowed from mesoscopic
systems may have much broader  domain of applications in the 
Euclidean QCD. Some preliminary analysis of the chiral 
disorder influenced by {\em external sources} seems to show, that this
 is indeed the case~\cite{USEXTERNAL}.
 The effects on disorder  of quark  chemical potential
resemble a complex electric Aharonov-Bohm effect, breaking
particle-antiparticle symmetry, accumulating two flux lines and
leading to rupture of "baryonic" quark-antiquark pair at
$\mu=M_{\pi}/2$.
This is observed in quenched lattice simulations and in 
random matrix models~\cite{STEPHANOV}.   
The phenomenon of persistence currents in disordered media finds an
analogy to diffusion of light quarks in the presence of several
Abelian Aharonov-Bohm fluxes. 
Some magnetic properties of chiral condensate can be explained by 
replacing the diffuson/cooperon  trajectories 
by the four-dimensional Landau orbits. Low energy theorems obtained 
in~\cite{SMILGASUSH} can be interpreted as {\it negative magnetoresistance}
of the QCD vacuum (quark condensate grows with the magnetic field).    
Low temperatures correspond to the replacement  of initial  cube
$L^4$ 
by asymmetric box $\beta L^3$, and the diffuson modes reproduce
the lowest temperature corrections for the chiral observables. 

An extremely interesting case corresponds to the situation, when
the critical temperature is reached. In the case of a  second 
order phase transition, pion wavefunction, susceptibility and the
condensate
undergo the following scaling
\be
Z_{\pi}=\Sigma^{\nu \eta/\beta},\,\,\,\,\,\,\,\,\,\,\,\,\,
\chi_{\pi}=\Sigma^{1 -\delta},    \,\,\,\,\,\,\,\,\,\,\,\,\,
\Sigma= m^{1/\delta}
\ee 
where I used the standard notation for the critical exponents.   
These effects have an obvious quantitative  effects on the diffuson
(pion),
modifying the return probability. In particular, the probability 
of return at the critical point tends to universal behavior 
$P(\tau)= (\tau_H/ \tau)^{1/\delta}$. 
This has to be contrasted with the vacuum result $P(\tau) \rightarrow 
\,\, const.$ in the ergodic regime and vacuum result $P(\tau) 
\rightarrow (\tau_c/\tau)^2$ in the diffusive regime. 
Finally, let me point that the concept of the return probability at
the Schwinger time allows to check  scenarios beyond the rigorous
theory of phase transitions. For metal-insulator  type transition, 
$P(\tau) \sim \tau^{1-\eta/4}$. The multifractal exponent $\eta$
estimated on the basis of chiral disorder is $\eta/4=0.057$~\cite{USEXTERNAL}.
 The smallness of $\eta$
makes the return probability  to look "diffusive" with $d=2$.  
 Finally, the asymmetry of the box may cause the appearance of two
 diffusion processes: "temporal" one and the "spatial" one. 
Since the spatial diffusion constant $D_s$ is small in 
the high temperature phase 
(typically, $D_s/D_t \sim 0.1$ at $T\sim 180$ MeV), 
 the asymmetry in conduction properties 
may cause a "percolation" from $d=4$ to $d=1$, with one-dimensional 
diffusive behavior $P(\tau) \sim 1/\sqrt{\tau}$. 
These speculations are interesting from the point of the view of
lattice spectral analysis in the vicinity of the critical temperature, 
since the return probability in the Schwinger time
can be explicitly expressed in terms of the eigenfunctions and
eigenvalues of the Dirac operator.

\subsection* { Threefold ways and lattice spectra }

Let me finish these lectures reminding the classification of the
global symmetries of the Dirac operator, hence the classification of
the chiral universality classes.

For complex quarks, the Dirac operator has no symmetries.
For real quarks (two colors), Dirac has an anti-unitary symmetry,
$[ \gamma_2\gamma_4 \sigma_2 K, iD\!\!\!\!/_F]=0$, where $K$ denotes
complex conjugation.  
For adjoint quarks Dirac operator  has another anti-unitary symmetry,
$[\gamma_2\gamma_4K,iD\!\!\!\!/_A]=0$, where 
subscripts $F,A$ denote the fundamental and adjoint representations, respectively.
This completes the list, forming the threefold way, how global
symmetries can be realized for the Dirac Hamiltonian~\cite{VERBTRIAD}:

{\bf I}.  For complex quarks with $N_f$ flavors , the pattern of spontaneous breakdown of the chiral
symmetry corresponds to $SU(N_f) \times SU(N_f) \rightarrow SU(N_f)$, and
the random matrix realization is obtained by filling the matrix with
complex numbers (chiral (denoted by $\chi$)
 Gaussian unitary ensemble ($\chi Gue$)). This is the case of 
the QCD, analyzed in these lectures.

{\bf II}. For real quarks with $N_f$ flavors,
 the pattern of spontaneous breakdown of the chiral
symmetry corresponds to $SU(2N_f)  \rightarrow Sp(2N_f)$, and
the random matrix realization is obtained by filling the matrix with
real  numbers ( chiral Gaussian orthogonal  ensemble ($\chi Goe$)). This is the case of 
QCD with two colors.

{\bf III}. For adjoint   quarks with even $N_f$,
 the pattern of spontaneous breakdown of the chiral
symmetry corresponds to $SU(N_f)  \rightarrow O(2N_f)$, and
the random matrix realization is obtained by filling the matrix with
quaternions  ( chiral Gaussian symplectic ensemble ($\chi Gspe$)).
 This is the case of 
QCD with adjoint fermions, for any number of colors.

For each of these ensembles, one can obtain exact spectral formulae.  
Historically, the first one was (\ref{micro}), obtained in 
\cite{VERBAARSCHOTZAHED}.
Technically, the complex case is the easiest one. 
  
This  triad is  distorted on the lattice, due to 
the known problems of incompatibility of having chiral fermions on the
discrete lattice with local action~\footnote 
{The original classification can be however reproduced using
nonlocal actions.}. 

For example, the additional symmetries of Kogut-Susskind Hamiltonian cause, 
that for two colors the lattice universality class is $\chi Gspe$, and
not, as expected from II, $\chi Goe$.  Historically, the microscopic
spectrum
for this ensemble was the first one explicitly 
confirmed by lattice, in~\cite{TILOSYMPL}.
The formula (\ref{micro}) 
was confirmed successfully on the  lattice ($N_f=0$ case) 
only recently~\cite{DAMGAARDGUE}.

Nowadays, there exist an  impressive list of theoretical 
predictions for several spectral (also higher-point) correlations
and plethora of lattice
evidence for various combinations of gauge groups, fermion
representations,
topological sectors, with quenched and  unquenched and "double scaled" 
determinants etc. We refer the curious reader to original literature. 
In all cases, the  agreement with RMM is excellent. 

There is also a growing interest in moving toward the diffusive
regime.
First,  the scaling (\ref{conductance}) was confirmed 
by lattice study~\cite{BERBENNI2}.  In the previous subsection we
 mentioned the measurements  of the crossover
between the ergodic and diffusive regime.
They are also first measurements on the chiral parameters, motivated
by the spectral properties of the Dirac operator.

\subsection* {Further reading} 
An excellent review on chiral dynamics  is~\cite{LEUTWYLER},
 where the reader
will find deeper justification of several statements 
made in  Part~1. 
The original arguments on diffusion in QCD , based on rather compact
paper \cite{PRLUS}  and the sequels \cite{USEXTERNAL},
 will become more obvious after 
reading condensed matter literature on this
 subject~\cite{BOOKDIFF,MONTAMBAUX}.
The consequences of Banks-Casher relation are discussed in 
\cite{LEUTWYLERSMILGA}.
Basic informations on the phenomenology of the instanton vacuum 
could be found in \cite{INSTANTONS}. 
Mehta' s book~\cite{MEHTA} is a classics in random matrix theory, 
for the reviews on the chiral ensembles we refer to \cite{ACTA,WETTIGVERB}.
Readers interested in broader aspects of RMM we refer to 
\cite{GUHR,GINSPARG}. 
I did not discuss here at all the schematic use of random models in
QCD
phenomenology, concentrating here on exact, quantitative predictions
in the ergodic window. Sample applications could be found in a review  
\cite{BLUE} and references therein.   
Finally, a guidebook to advanced details (supersymmetric technique,
 replicas etc. )
of the partially  quenched perturbation techniques  could be found in
recent reviews ~\cite{WETTIGVERB,DAMGAARDVERLAST} and references therein.

\section*{Conclusions}
In these lectures, I tried to demonstrate in {\it a physical way} how 
much the concepts of the spontaneous breakdown of the chiral symmetry
share with the concepts of the disorder in mesoscopic systems.
This should comply  with the introductory character of these lectures.
I perhaps omitted several important works to quote, and I apologize
for
this. The main idea of these lectures was to introduce a non-expert
to the fascinating field of chiral disorder in the QCD. 
I therefore sacrificed  several advanced technical details 
and  mathematical aspects, which are often
necessary
to prove  the analogies presented here  in a strict way.
Luckily, there exist reviews, some quoted above,
 which  exhaustively can guide the
 reader
through the technicalities of random matrices and various versions of 
chiral perturbation theories adapted for finite volume systems.
They also   
provide the more complete bibliography.

I pointed, that the ergodic (universal) domain, 
 is to large extent understood and 
documented in an impressive way. Universality of the ergodic regime 
helped lattice QCD to check to what extend chiral properties
(including the global features e.g. zero modes) of the
quarks are reflected by simulation. They also allowed to quantify the 
discrepancies, 
in the light of the exact analytical predictions of the RMM. 
I tried  to emphasize, how   
challenging is to look at the spectra of  the diffusive regime.
The lattice analogs of the diffuson theories
open a way to systematic extraction of the effective parameters
of the chiral Lagrangian, following step by step the Weinberg expansion.
(The first  example is a pion decay constant).
It is also possible to study the  dependence of these constants
of the chiral Lagrangian  
on the external
parameters and it is important  to see what happens to them 
when the critical 
regime is approached. 
It is also plausible, that the spectral analysis in the
diffusive regime  may serve as an important tool for studying the 
nonperturbative nature
of the gluon fields and their quantitative rearrangements in the
vicinity
of the phase transition, hopefully shading some more light on the 
nature of the lumps of the "color dirt".     

A more general message, which the reader should infer from these
lectures and from several  other lectures  at this school is,
 how broadly and how
fruitfully 
the theory
of strong interactions can borrow from the concepts and ideas 
of  the condensed matter theory.   

\begin{acknowledgments}
The philosophy of these lectures (if one is permitted to use such an
expression) is greatly influenced by  my present collaborators on
the chiral disorder:  
Romuald Janik, Jurek Jurkiewicz,  Gabor Papp and Ismail Zahed. 
Some early studies of chiral disorder and  on random
matrices in instanton models, which triggered my interest in this
subject,  were done in collaboration with Reinhard Alkofer, 
Jac Verbaarschot and Ismail Zahed. It is a pleasure to notice, that these
and other early attempts~\cite{EARLY} 
  have come logically  full  circle, with so powerful
applications and predictions.  
    
  I would like to thank  Jean-Paul Blaizot,
Edmond Iancu, Andrei Leonidov  
 and Larry Mc Lerran  for an  invitation  to
deliver these lectures and for creating 
 an imaginative and inspiring  atmosphere of
the 2001 Carg\'{e}se Summer School.  
 
 I  would like to  thank also 
Wolfgang Noerenberg,  Hans Feldmeier,  Bengt Friman, Joern Knoll, 
Volker Koch and Matthias Lutz   
for discussions and for providing the excellent opportunity to
 complete these lectures during my visit at GSI.

\end{acknowledgments}

\end{document}